\documentclass[nolinenumbers]{aastex631}

\usepackage{gensymb}
\usepackage{textcomp}
\usepackage{amsmath}
\usepackage{booktabs}
\usepackage{hyperref}
\usepackage{xcolor}
\newcommand{\G}[1]{\textcolor{black}{#1}} 

\hypersetup{
    colorlinks=true,
    urlcolor=blue
}

\begin{document}

\title{A High-Accuracy Alignment Approach for Solar Images of Different Wavelengths}

\author[0009-0007-6584-465X]{Yun Wang}
\affiliation{Yunnan Observatories, Chinese Academy of Sciences, Kunming 650216, China \\}
\affiliation{University of Chinese Academy of Sciences, Beijing 101408, China \\}

\correspondingauthor{ KaiFan Ji}
\email{jkf@ynao.ac.cn}
\author[0000-0001-8950-3875]{Kaifan Ji }
\affiliation{Yunnan Observatories, Chinese Academy of Sciences, Kunming 650216, China \\}

\author[0000-0001-7575-5449]{Zhenyu Jin}
\affiliation{Yunnan Observatories, Chinese Academy of Sciences, Kunming 650216, China \\}
\affiliation{Yunnan Key Laboratory of Solar Physics and Space Science, 650216, China \\}

\author[0000-0003-2714-6811]{Hui Liu}
\affiliation{Yunnan Observatories, Chinese Academy of Sciences, Kunming 650216, China \\}



\begin{abstract}

Image alignment plays a crucial role in solar physics research, primarily involving translation, rotation, and scaling. \G{The different wavelength images of the chromosphere and transition region have structural complexity and differences in similarity, which poses a challenge to their alignment.} Therefore, a novel alignment approach based on dense optical flow (OF) and the RANSAC algorithm is proposed in this paper. \G{It takes the OF vectors of similar regions between images to be used as feature points for matching. Then, it calculates scaling, rotation, and translation.} The study selects three wavelengths for two groups of alignment experiments: the 304 {\AA} of the Atmospheric Imaging Assembly (AIA), the 1216 {\AA} of the Solar Disk Imager (SDI), and the 465 {\AA} of the Solar Upper Transition Region Imager (SUTRI). Two methods are used to evaluate alignment accuracy: Monte Carlo simulation and Uncertainty Analysis Based on the Jacobian Matrix (UABJM). \G{The evaluation results indicate that this approach achieves sub-pixel accuracy in the alignment of AIA 304 {\AA} and SDI 1216 {\AA}, while demonstrating higher accuracy in the alignment of AIA 304 {\AA} and SUTRI 465 {\AA}, which have greater similarity.}

\end{abstract}



\section{Introduction} \label{sec:intro}

The sun emits electromagnetic radiation across various wavelengths, including infrared, visible light, ultraviolet, extreme ultraviolet, and X-rays. The observation of these different wavelengths provides insight into the physical processes occurring in the solar atmosphere at varying heights and temperatures. Therefore, multi-wavelength observations provide comprehensive and three-dimensional information for studying solar activity. Multi-wavelength observation is an important method for empirical solar physics research. Many solar activity phenomena exhibit different observational characteristics across different radiation wavelengths, with varying brightness and spatial forms. Image alignment at a single wavelength facilitates observation and study of the sun's evolution at that specific wavelength. Conversely, performing image alignment across multiple wavelengths facilitates a comprehensive analysis of these activity phenomena, thereby helping to uncover the patterns of solar activity.

A telescope pointing and tracking system can be used for single-wavelength image stabilization \citep{Staiger2013}. This makes it possible to acquire information like heliocentric coordinates and align single-wavelength images. However, even if the telescope is designed to have sub-arcsecond pointing and tracking accuracy, problems such as bending of the optical support system due to its structure and thermal jitter can lead to inaccuracies during the observation process. \citet{Shimizu2007} utilized two Ultra Fine Sun Sensors (UFSS-A and UFSS-B) to detect satellite jitter. In fact, not only can jitter measurements be realized by hardware, but also various algorithms can be used to detect the jitter problem. \citet{Orange2014},for example, performed pointing jitter measurements on the Helioseismic and Magnetic Imager (HMI) and AIA on the Solar Dynamics Observatory (SDO) using a mutual correlation algorithm. In contrast to the alignment of single-wavelength images, multi-wavelength image alignment is the alignment of solar images from different wavelengths. These images usually originate from different wavelengths of the same observing instrument or from different wavelengths of different observing instruments. There may also be small offsets and pixel size differences between images taken at different wavelengths by the same instrument \citep{Guglielmino2010}. \citet{Shimizu2007} evaluated the internal offsets and size differences of the broadband filter imager of the Solar Optical Telescope (SOT). The alignment of images across different wavelengths poses a significant challenge due to the varied criteria associated with different wavelengths and instruments. This challenge is further compounded by the presence of scale, rotation, and translation differences, which can arise from instrument differences, wavelength differences, and disparities in image processing techniques. Therefore, various algorithms need to be developed to realize image alignment between different wavelengths.

Typically, image alignment involves the estimation of translation, rotation, and scaling. With the increasing demand for fine information in solar physics research, the accuracy requirements of image alignment have increased. \G{At present, two primary classical methods are employed for solar image alignment.}
One is a region-based statistical method, which maximizes the correlation between images through the statistical information of image regions to achieve alignment. Cross-correlation (CC) and phase correlation (PC) are two common statistical alignment methods. \citet{Kuehner2010} achieved multi-wavelength alignment on HINODE/SOT by CC algorithm, and \citet{Berkebile2009} also used CC algorithm to achieve the image alignment of Dutch open telescope (DOT) and the transition region and coronal explorer (TRACE). However, CC performs well in sub-pixel accurate translation transformations but has difficulty in scale and rotation transformations. Conversely, the PC algorithm has been developed to achieve rotation and scale transformations between images using Fourier transform, Polar Coordinate transform, and Logarithmic transform \citep{Reddy1996}. \citet{Druckm2009} achieved the alignment of coronal images during a total solar eclipse by measuring the translations, rotations, and scale factors between images with the PC algorithm. The IPC algorithm is an extension of the PC algorithm that utilizes the differential evolution algorithm to optimize the parameters and improve the effectiveness and accuracy of the algorithm \citep{Hrazdra2020}.
Another approach is the feature-based matching method, which utilizes salient regions, lines, or points in the image as distinct reference features. \citet{Lowe1999,Lowe2004} designed and developed the scale invariant feature transform (SIFT) method by combining the steps of feature point detection, vector generation, and matching search. In the solar photosphere, the spot features are obvious, thus prompting the application of the SIFT method in the field of astronomy \citep{Yue2015}. \citet{Yangpan2018} employed this method to align and localize the local solar magnetic field from the Huairou Solar Observatory (HSO) with full-disk solar magnetic field images from SDO/HMI. Later, \citet{Ji2019} employed SIFT to align HMI, GONG, and AIA 304 {\AA} data with TiO and H$\alpha$ wavelength images acquired by NVST. In addition, SIFT was utilized to conduct a search for solar active regions \citep{Jiang2022}.

Optical flow (OF) represents a significant research direction within the domain of computer vision, with applications including target recognition and tracking. Recently, OF methods have also been applied to the alignment of solar images. \citet{Cai2022} utilized the OF algorithm to align the H$\alpha$ data of the NVST and evaluated its performance accuracy with raster images obtained from the Fast Imaging Solar Spectrometer (FISS) run by the GST. \G{Moreover, the accuracy of the method is higher than the CC algorithm.} \citet{Yang2022} used OF and SIFT algorithms to align the data from GST. However, both of them only utilized OF for translational direction. Also, the OF method can be used for high-resolution solar image reconstruction \citep{Liu2022}.

\begin{figure}[ht!]
\plotone{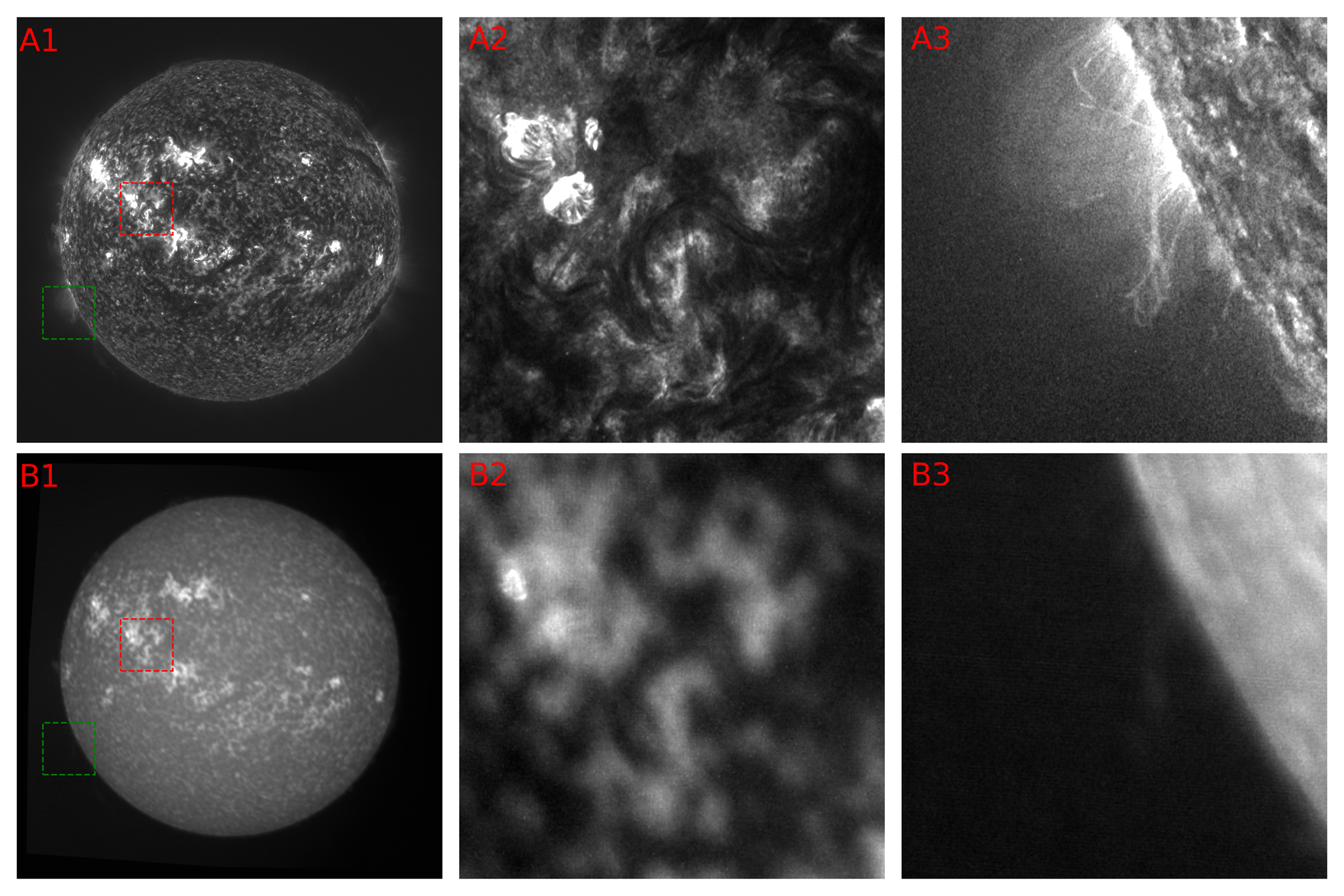}
\caption{ \G{Comparison image between AIA 304 {\AA} and SDI 1216 {\AA}. A1 is AIA 304 {\AA}, B1 is SDI 1216 {\AA}. A2 and B2 are the comparison of their explosion areas, while A3 and B3 are the comparison of their edge areas.} 
\label{fig:data}}
\end{figure}

The distinct observational characteristics exhibited by different wavelengths are attributable to the varied solar atmospheres observed. To realize the alignment of different wavelengths of solar images, it is necessary to require similar structures between these wavelengths. However, in scenarios where the similarity between images is low, particularly when only a portion of the similar structure is present, achieving an accurate alignment between images can be a formidable challenge. \G{Currently, the correlation algorithms are reliant on the similarity between images. If there are only partially similar structures between wavelengths, the accuracy of the correlation algorithm is decreases.} For feature point matching algorithms, the higher the number of feature points, the higher the alignment accuracy usually is. In practical applications, the number of feature points in solar images is often small, and there is too much manual intervention, making it difficult to improve the alignment accuracy. \G{When aligning images from the SDI 1216 {\AA} and AIA 304 {\AA} channels, the two datasets exhibit significant differences in spatial resolution, leading to inherently low global similarity. And regions and edge boundaries demonstrate limited structural correspondence (Figure \ref{fig:data}). We use the \citet{Bhattacharyya1943OnAM} coefficient to quantify the similarity between the images. A subsequent comparison of the intensity histograms of the solar images yielded a Bhattacharyya coefficient of 0.58 for SDI 1216 {\AA} and AIA 304 {\AA} and 0.79 for SUTRI 465 {\AA} and AIA 304 {\AA}.} Therefore, the alignment effect of these two methods is not ideal.

In this paper, a novel solar image alignment approach based on dense OF and the RANSAC algorithm is proposed. It is able to realize region-based matching by using the partially similar structure between images on solar images where feature points are difficult to find. Then, it detects information such  as translation, rotation, and scale. The paper is organized as follows: Section \ref{sec:data} describes the solar image data used for alignment; Section \ref{sec:Alignment Method} details the alignment method using the example of SDI 1216 {\AA} and AIA 304 {\AA} alignment; Section \ref{sec:Alignment accuracy evaluation and Results} evaluates the alignment accuracy of the method and shows the results; and conclusions and discussions are given in Section \ref{sec:V.Conclusion and Discussion}.

\section{Data} \label{sec:data}

Three data groups are selected for analysis: AIA 304 {\AA}, SDI 1216 {\AA}, and SUTRI 465 {\AA}. And we can validate the proposed approach through these data. The three selected data groups are from neighboring regions of the solar atmosphere in adjacent time periods and have partially similar structural features. Among them, AIA 304 {\AA} is primarily employed for the observation of the chromosphere and the transition region, SDI 1216 {\AA} is utilized for the observation of the chromosphere-to-corona region of the Sun, and SUTRI 465 {\AA} focuses on the upper transition region of the Sun. In this image alignment experiment, two groups of data are selected for the purpose of aligning different wavelengths: SUTRI 465 {\AA} and AIA 304 {\AA} data on November 14, 2022, and SDI 1216 {\AA} and AIA 304 {\AA} data on January 31, 2024.

The Atmospheric Imaging Assembly (AIA) was launched on February 11, 2010 from the Solar Dynamics Observatory (SDO) \citep{Pesnell2012}. The AIA is capable of observing ten different wavelengths, including seven extreme ultraviolet, two ultraviolet, and one visible wavelength. In in-orbit observations, AIA 304 {\AA} has a spatial resolution of 1.5\(''\) and a temporal resolution of 12 s, generating images of 4096\(\times\)4096 pixels\(^{2}\). The image scale is 0.6\(''\) pixel\(^{-1}\) \citep{Lemen2012}.

The Lyman-alpha (Ly$\alpha$) Solar Telescope (LST) is one of the payloads on board the Advanced Space-based Solar Observatory (ASO-S), which was successfully launched on October 8, 2022 \citep{Gan2023}. The Solar Disk Imager (SDI) is an instrument onboard the LST with an operating spectrum of the Ly$\alpha$ line (1216 {\AA}) \citep{Chen2019}. In the in-orbit observations, SDI has a spatial resolution of approximately 9.5\(''\) \citep{Chen20024} and a temporal resolution of 10 s, generating images of 4608\(\times\)4608 pixels\(^{2}\). The image scale is 0.5\(''\) pixel\(^{-1}\) \citep{Li2019}.

The Solar Upper Transition Region Imager (SUTRI) was carried on board the SATech-01 satellite of the Chinese Academy of Sciences and was successfully launched on July 22, 2022, with an orbital period of 96 min \citep{Zhang2024}. SUTRI operates at the spectral line of Ne VII 465 {\AA} and is mainly formed in the transition region in the solar atmosphere above 0.5 MK degrees. In the in-orbit observations, SUTRI has a spatial resolution of 8\(''\) and a temporal resolution of 30 s, generating images of 2048\(\times\)2048 pixels\(^{2}\). The image scale is 1.229\(''\) pixel\(^{-1}\) \citep{Bai2023}.

One image from each of the above three data groups is selected as a reference for the simulation of the alignment experiment. The generation of three test groups is achieved through the presetting of four parameters: scale, rotation angle, x-direction displacement, and y-direction displacement. Each test group contains 1000 randomly generated images that have undergone similarity transformation. The alignment evaluation of the single-wavelength simulated data is performed in Section 4 through three test groups.

\section{Alignment Method} \label{sec:Alignment Method}

Given the differences in the solar atmospheric regions observed by AIA 304 {\AA} and SDI 1216 {\AA}, we aim to align the two images by focusing on the similar regions that are common to both. \G{It is evident that these two wavelengths contain similar structures within the quiet region of the sun.} Utilizing the dense OF method facilitates the extraction of these similar structures. And due to the inherent limitations of the OF algorithm, its capacity to process a wide range of movement is constrained. Consequently, it is necessary to first perform a coarse alignment of the image prior to calculating the OF vectors. \G{The utilization of the dense OF algorithm gives rise to mismatched OF vectors. These erroneous vectors are observed in active solar regions, such as flares, as well as in the sun's limb, which are dissimilar regions.} Therefore, it is necessary to eliminate these dissimilar regions with the help of the RANSAC algorithm and retain only the similar regions between images. Concurrently, the OF within these regions is employed as the acquired feature points to fit the similarity transformation model. Subsequently, the translation, rotation, and scale parameters between the images are derived.

\begin{figure}[ht!]
\plotone{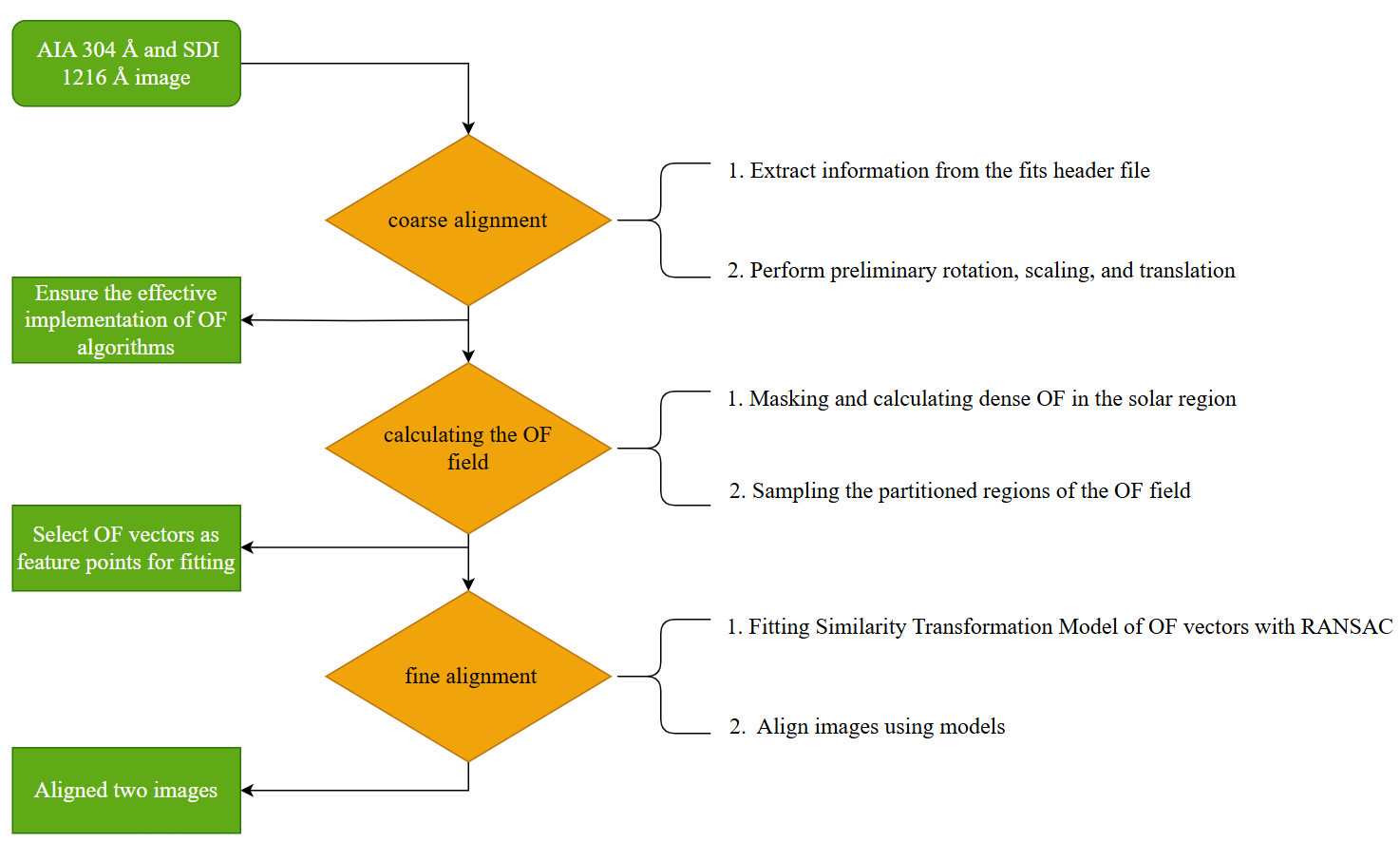}
\caption{\G{Alignment flowchart for AIA 304 {\AA} and SDI 1216 {\AA}. }
\label{fig:flowchart}}
\end{figure}

In summary, the alignment method proposed in this paper consists of three main steps: First, the process of coarse alignment is executed by employing the FITS header file data of the two solar images; Second, the OF field of the two images is calculated following the coarse alignment; Lastly, the OF vectors within the OF field are used as feature points to fit the similarity transformation model using RANSAC. This process enables the derivation of the similarity transformation matrix and the realization of the fine alignment of the images. \G{Figure \ref{fig:flowchart} shows the specific registration process of AIA 304 {\AA} and SDI 1216 {\AA}.}

\subsection{Coarse Alignment} \label{subsec:Coarse Alignment}

\G{Initially, a preliminary alignment of the images must be conducted to ascertain that the geometric transformations between the two solar images are small.} This is necessary to facilitate the effective implementation of the OF algorithm. Typically, the FITS header file of an image contains essential information such as the rotation angle, the pixel scale, and the sun center position. Leveraging this information, we are able to perform a coarse alignment of solar images in different wavelengths using similarity transformation. However, given the differences in observation equipment and image processing methods, this alignment may still lead to image errors in practice. \G{Thus, a subsequent fine alignment is required.}

\subsection{Calculate the OF field} \label{subsec:Calculate the OF field}

Optical flow, as an important research area in the field of computer vision, is the instantaneous velocity that describes the motion of pixels of a spatially moving object on the observation imaging plane \citep{Horn1981}. The OF method calculates the OF field on an image. The field consists of a large number of OF vectors. The OF method is employed in the domains of object recognition and tracking. It utilizes the variation of pixels in an image sequence in the time domain and the correlation between adjacent frames to find the correspondence that exists between the previous frame and the current frame. This method enables the calculation of motion information between adjacent frames, facilitating the analysis of object motion.

The OF method has three basic assumptions: (1) constant brightness: the brightness of the same target does not change when it moves between frames; (2) time consistency: changes in time will not cause drastic changes in the target position, and the displacement between neighboring frames should be relatively small; (3) spatial consistency: object motion in an image is typically smooth, with neighboring pixel points exhibiting similar velocities and orientations.The aforementioned assumptions can be expressed as follows:
\begin{equation}
I(x,y,t) = I(x+\Delta x ,y+\Delta y,t + \Delta t),
\end{equation}
where $I$ denotes the pixel intensity at $(x,y)$ in the first frame. After a time interval of \(\Delta t\), it moves the displacement (\(\Delta x\),\(\Delta y\)) to the next frame. Due to the small magnitude of the motion, the right-hand side of the equation is obtained by performing a first-order Taylor expansion and neglecting the higher terms:
\begin{equation}
\frac{\partial I}{\partial x} \frac{\mathrm{d} x}{\mathrm{d} t}  +\frac{\partial I}{\partial y} \frac{\mathrm{d} y}{\mathrm{d} t} +\frac{\partial I}{\partial t}=0 ,
\end{equation}
where $\frac{\partial I}{\partial x}$ and $\frac{\partial I}{\partial y}$ are the spatial gradients of the image and $\frac{\partial I}{\partial t}$ is the gradient in the temporal direction. And $\frac{\mathrm{d} x}{\mathrm{d} t}$ and $\frac{\mathrm{d} y}{\mathrm{d} t}$, are the OF vector and the unknown quantities to be solved.

At present, there are many algorithms and theories to calculate the OF field. The Gunnar farneback algorithm \citep{farneback2003} is a type of dense OF that calculates the motion information of pixels individually. It generates a Gaussian pyramid of images with different resolutions for multi-resolution image search. While this algorithm is more time-consuming than other OF methods (such as sparse optical flow), it is capable of achieving high accuracy for images with complex structures. For this reason, it was selected to calculate the motion information for each pixel in the solar image. Although the OF vector itself describes the instantaneous velocity of pixels between images, it can be approximated as equal to the pixel displacement under certain circumstances. The image data in this paper is calculated using the Gunnar farneback algorithm. The resulting OF field is the relative displacement field of the image pixels.

It is necessary to implement image masking and region sampling before acquiring the OF field for executing RANSAC. In essence, the calculation of the OF field necessitates a concentration on the solar region alone, obviating the need for the calculation of the entire image. Therefore, the mask can be employed to calculate only the sun component. Given that the sun is already centered in the image during the coarse alignment process, the mask can be utilized to remove the sun's edges and the subsequent regions. This approach enables the reduction of both the calculated burden and the unreliable sun edge regions. Subsequent to the calculation of the dense OF for the sun region, the sampling operation is performed by dividing the sun region. The Gunnar Farneback OF algorithm generates a dense OF field, comprising the OF vector at each pixel point. The selection of the median OF vector within each region is achieved by dividing the regions for sampling. This approach ensures the homogeneity of the solar surface OF field during the subsequent fitting process. Thus, it avoids overfitting of specific regions, which would otherwise lead to a shift in the overall fitting results. By doing the above, we can obtain the OF field used for the RANSAC operation, as shown in the middle image of Figure \ref{fig:ransac}.

\subsection{Fine Alignment} \label{subsec:Fine Alignment}

The Random Sample Consensus (RANSAC) algorithm \citep{Fischler1981} estimates the parameters of a mathematical model from a group of observed data containing outliers. This estimation is achieved through an iterative approach. Compared to the least squares method, it incorporates the concept of rejecting outlier data. Consequently, it facilitates the expeditious and precise identification of data samples that contain erroneous data.

The OF field obtained is not reliable in the solar active regions or in the solar limbic region due to the limitations of the assumptions of the OF algorithm. Accordingly, the OF field calculated by the Gunnar Farneback algorithm contains a number of outlier points. The outliers can be quickly and accurately screened out using the RANSAC algorithm, and the model is fitted with the similarity transformation. The similarity transformation matrix possesses four degrees of freedom, namely the scaling factor, rotation, x-direction displacement, and y-direction displacement. The specific model is as follows:
\begin{equation}
\begin{bmatrix}
 x^{\prime } \\
 y^{\prime }\\
 1
\end{bmatrix}=
\begin{bmatrix}
  s \cos \beta  &  -s \sin \beta& \mathrm{d}x \\
  s \sin \beta&  s \cos \beta& \mathrm{d}y\\
  0&  0&1
\end{bmatrix}
\begin{bmatrix}
 x\\
 y\\
1
\end{bmatrix},
\end{equation}
where $(x, y)$ is the pixel position of the first image$(x^{\prime}, y^{\prime})$, is the corresponding pixel position of the next image, $s$ is the scaling factor, $\beta$ is the rotation angle, and $\mathrm{d}x$ followed by $\mathrm{d}y$ are the displacements in the x and y directions.

\begin{figure}[ht!]
\plotone{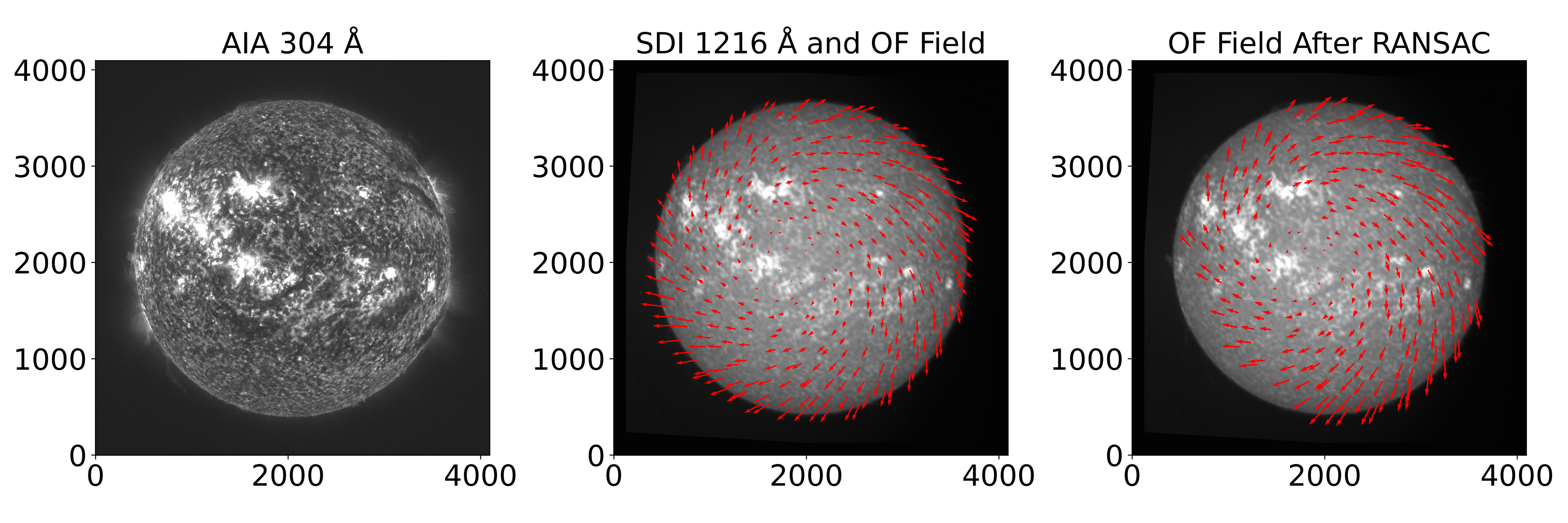}
\caption{Results of the OF field between images are shown. On the left is the AIA 304 {\AA} image used as a reference image; in the center is the SDI 1216 {\AA} image and the regionally sampled OF field; and on the right is the SDI 1216 {\AA} image and the effective OF field after RANSAC. 
\label{fig:ransac}}
\end{figure}

The OF field, as determined by the Gunnar Farneback algorithm, comprises both pixel coordinates and displacements. This provides the corresponding pixel positions within the similarity transformation matrix.  The RANSAC algorithm can then be utilized to filter out outliers to preserve the OF field with similar structure between images. The right part of Figure \ref{fig:ransac} shows the OF field after the RANSAC algorithm filters out the outliers. \G{The figure shows that the filtered OF vectors remove part of the solar limb where there are significant differences.} Moreover, the OF vectors are reduced in active regions such as flares and are abundant in quiet regions. This improves the accuracy of the alignment in subsequent fits. The distribution of the OF vectors indicates the presence of rotation between images. Utilizing the OF vectors as  feature points, we fit the similarity transformation model and thus solve for the four parameters. Then, the four resolved parameters can be utilized to perform the similarity transformation on the image to be aligned, thereby ensuring fine alignment with the reference image.

\section{Alignment accuracy evaluation and Results} \label{sec:Alignment accuracy evaluation and Results}

\G{We select two approaches for evaluation: Monte Carlo simulation and UABJM, to validate the alignment accuracy of our method. The three datasets in this experiment are obtained from different instruments with varying wavelengths. They inherently lack precise alignment relationships. So it becomes challenging to directly assess cross-band alignment accuracy. We therefore conduct preliminary evaluations using Monte Carlo simulation on single-wavelength data. Subsequently, we introduce the CC algorithm to compare alignment accuracy in translational dimensions. To demonstrate our method's noise robustness, we perform additional validation by incorporating Gaussian noise into solar images.
The UABJM method is employed for accuracy assessment for cross-wavelength alignment verification. The validity of UABJM for single-wavelength evaluation is first confirmed through comparison with Monte Carlo simulation results. Following this confirmation, we extend the UABJM methodology to evaluate different wavelength alignment accuracy.}

\subsection{Monte Carlo simulation} \label{subsec:Monte Carlo simulation}

Monte Carlo simulation estimates target statistics or expected values by generating a large number of random samples and analyzing their computational outcomes. \G{In our experiments, we establish accurate alignment relationships between simulated test datasets and original images through pre-defining four critical parameters (scale factor, rotation, x-translation, and y-translation). This framework enable systematic Monte Carlo simulations for three distinct single-wavelength datasets. The implementation procedure consists of four key phases:
First, to eliminate interference from solar rotation and active region flares, we select one high-quality observational image as the reference template. Subsequently, we randomly generate 1000 parameter sets (comprising floating-point values for the four transformation parameters) using uniform probability distributions. These randomized parameters are applied to perform similarity transformations on the original image, thereby creating comprehensive test datasets.
Following this implementation, we employ our methods to calculate the measured parameters. Systematic comparison between these measured values and true parameters yield alignment residuals. Finally, we quantify the alignment accuracy by calculating the root mean square error (RMSE) across all 1000 residual sets, with detailed results presented in Table \ref{tab:rmse1}.}

\G{The Gunnar Farneback algorithm for calculating the OF field needs to provide a pixel window to detect pixel motion information in our approach.} The size of this window affects the simulation accuracy, while the simulation range of the four parameters affects the window size. To ensure equivalent conditions for the evaluation of simulated data, we provide the same arcsec window (about 63\(''\)) for all image data. \G{Our approach demonstrates high accuracy in single-wavelength simulated data. Table \ref{tab:rmse1} shows that all wavelengths have scale errors $< 5 e^{-6} $, rotation errors $< $1\(''\), and displacement direction errors  $< $0.01 pixels.}

\begin{figure}[ht!]
\plotone{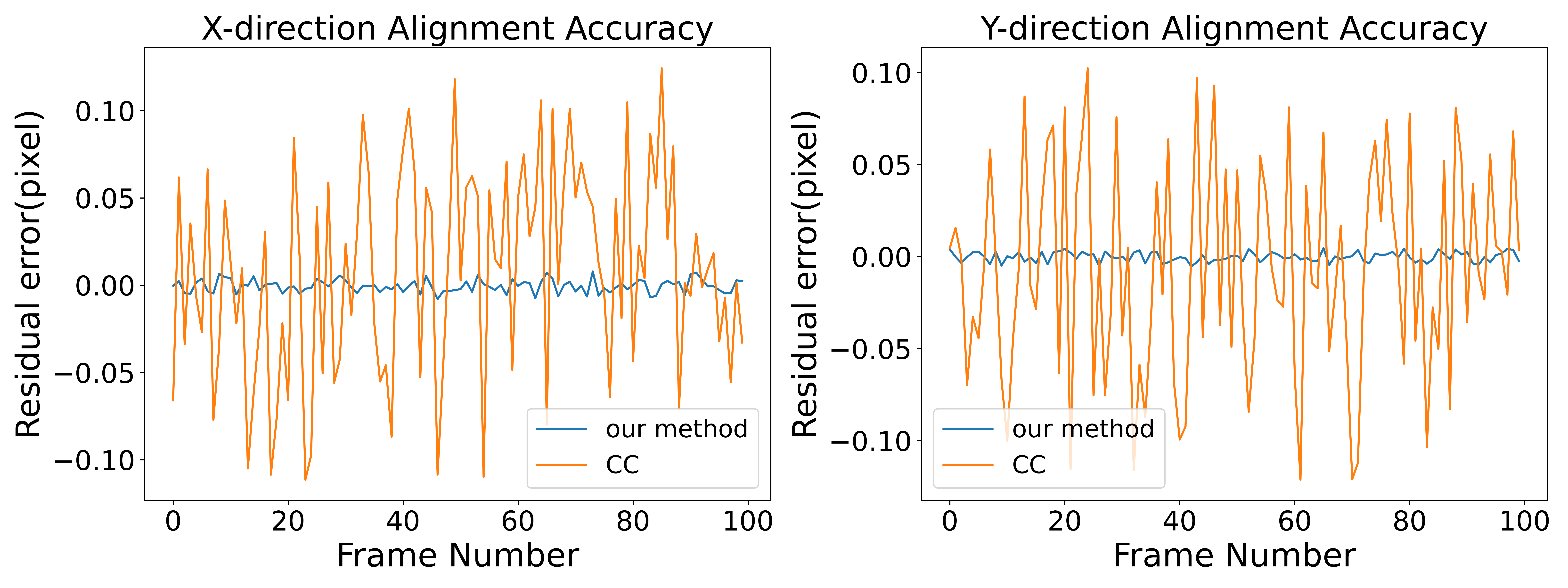}
\caption {\G{The alignment accuracy of 100 images using two different methods. The residual errors of x- and y-directions are plotted in two panels. Simulation range: x-direction -10 : 10 pixels; y-direction -10 : 10 pixels. The AIA 304 {\AA} image is selected here for comparison testing.}
\label{fig:dx-dy}}
\end{figure}

\begin{figure}[ht!]
\plotone{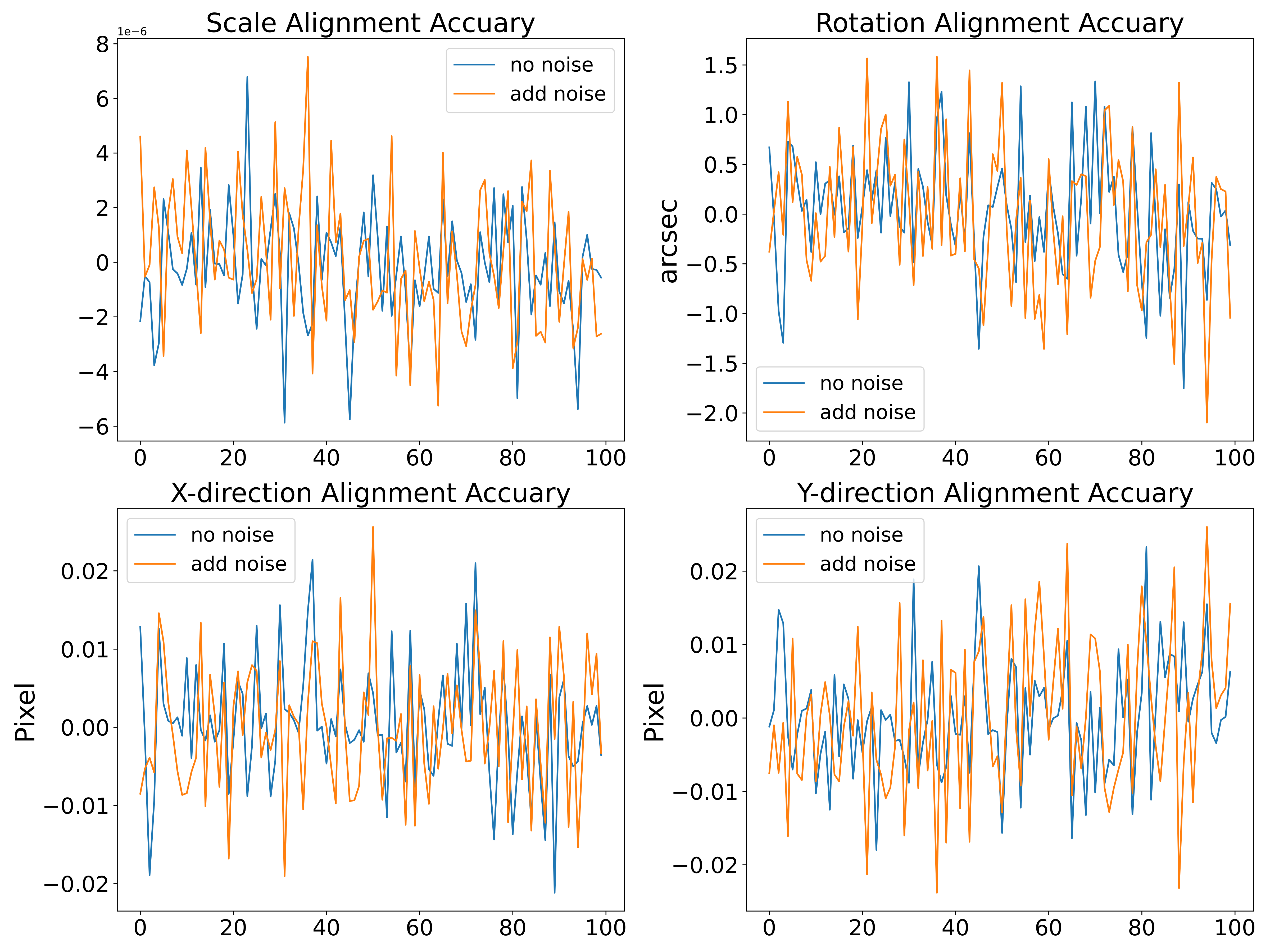}
\caption {\G{Comparison of the accuracy of four parameters with and without noise. The AIA 304 {\AA} image is selected here for noise testing.}
\label{fig:noise}}
\end{figure}

\G{At the same time, we conduct a comparative analysis with the CC algorithm targeting translational accuracy. By configuring predefined displacement parameters in both x- and y-directional axes, we perform 100 independent measurement trials for our approach and the CC algorithm. As demonstrated in Figure \ref{fig:dx-dy}, the comparison of alignment accuracy reveals the superior performance of our proposed method over the CC algorithm. The RMSE of the CC algorithm for x-axis and y-axis alignment accuracy is 0.0597 and 0.0582, respectively. And the RMSE of our approach for x-axis and y-axis alignment accuracy is 0.0037 and 0.0026, respectively.
Furthermore, we conduct a systematic noise robustness evaluation of our method. Gaussian noise (zero-mean; standard deviation equivalent to three times the background noise standard deviation) is introduced to the original solar image, followed by 100 simulated measurements. Figure \ref{fig:noise} presents a comparative visualization of measurement outcomes for AIA 304 {\AA} images under no-noise and add-noise conditions. Quantitative analysis reveals that the proposed method maintains measurement stability, with  a slow degradation of accuracy even under strong noise contamination. As detailed in Table \ref{tab:rmse1}, SUTRI 465 {\AA} exhibits error amplification. This discrepancy stems from the fact that the data from SUTRI have higher background noise and lower image quality than the remaining two groups.}

\begin{table*}[t]
    \centering
    \caption{RMSE of residual in the Monte Carlo simulation.}
    \label{tab:rmse1}
    \begin{tabular*}{\textwidth}{@{\extracolsep{\fill}}lccccr}
    \toprule
        \textbf{RMSE} & \textbf{} & \textbf{Scale} & \textbf{Rotation} & \textbf{x-direction} & \textbf{y-direction} \\ 
        \textbf{} &  \textbf{} &\textbf{$ e^{-6} $} & \textbf{arcsec} & \G{\textbf{pixel}} & \G{\textbf{pixel}}\\
        \midrule
        AIA 304 {\AA} & & 1.9017 & 0.6046 & \G{0.0093} & \G{0.0063} \\ 
            & \G{Add noise} & \G{2.4336} & \G{0.7052} & \G{0.0082} &\G{0.0102} \\
        SDI 1216 {\AA} & &  2.3404 & 0.4007 & \G{0.0088} & \G{0.0092} \\
        & \G{Add noise} & \G{2.0870} & \G{0.7580} & \G{0.0107} &\G{0.0095} \\
        SUTRI 465 {\AA} & &  4.5216 & 0.5263 & \G{0.0055} & \G{0.0056} \\
          &\G{Add noise} & \G{8.9630} & \G{1.6681} & \G{0.0110} & \G{0.0096} \\
    \bottomrule
    \end{tabular*}
    \tablecomments{Simulation range: scale 0.995 : 1.005; rotation -0.01 : 0.01 rad; x-direction -5 : 5 pixels; y-direction -5 : 5 pixels.}
\end{table*}

\subsection{Uncertainty Analysis Based on the Jacobian Matrix (UABJM)} \label{subsec:Uncertainty Analysis Based on the Jacobian Matrix (UABJM)}

The Monte Carlo simulation in Section 4.1 can only evaluate the alignment accuracy in a single wavelength and cannot evaluate the accuracy in different wavelengths. Consequently, an alternative error evaluation method is necessary to ascertain the accuracy of the alignment across different wavelengths. Considering the fitting of the similarity transformation matrix using RANSAC, then we can evaluate the accuracy of this alignment approach with the Jacobian matrix.

The specific steps of the UABJM are as follows: First, for the similarity transformation matrix model, the Jacobian matrix is calculated. Second, for the model fitted by RANSAC, we calculated the variance of its residuals. The expression is as follows:
\begin{equation}
S(\theta )=\sum_{i=1}^{n} \left [ Y_{i} -f(X_{i},\theta  ) \right ] ^{2} ,
\end{equation}
\begin{equation}
\sigma ^{2} =\frac{S(\theta )}{n-p} ,
\end{equation}
where \(X_{i}\) and \(Y_{i}\) denote the coordinates of the two images sought by the OF, $f$ denotes the fitting model, $\theta$ is the four parameters, $n$ is the sample size, $p$ is the number of parameters, here $p = 4$, $S$ is the residual sum of squares, and $\sigma ^{2}$ is the variance of the residuals. We can then estimate its covariance matrix with the following expression:
\begin{equation}
Cov(\theta )=(J^{T} J)^{-1}\sigma ^{2} ,
\end{equation}
where $J$ is the Jacobian matrix and $Cov$ is the covariance matrix, the diagonal arithmetic square root of which is referred to as the standard error estimates of the four parameters. Finally, the standard errors of the four parameters are estimated using the covariance matrix to serve as the alignment accuracy.

\begin{figure}[ht!]
\plotone{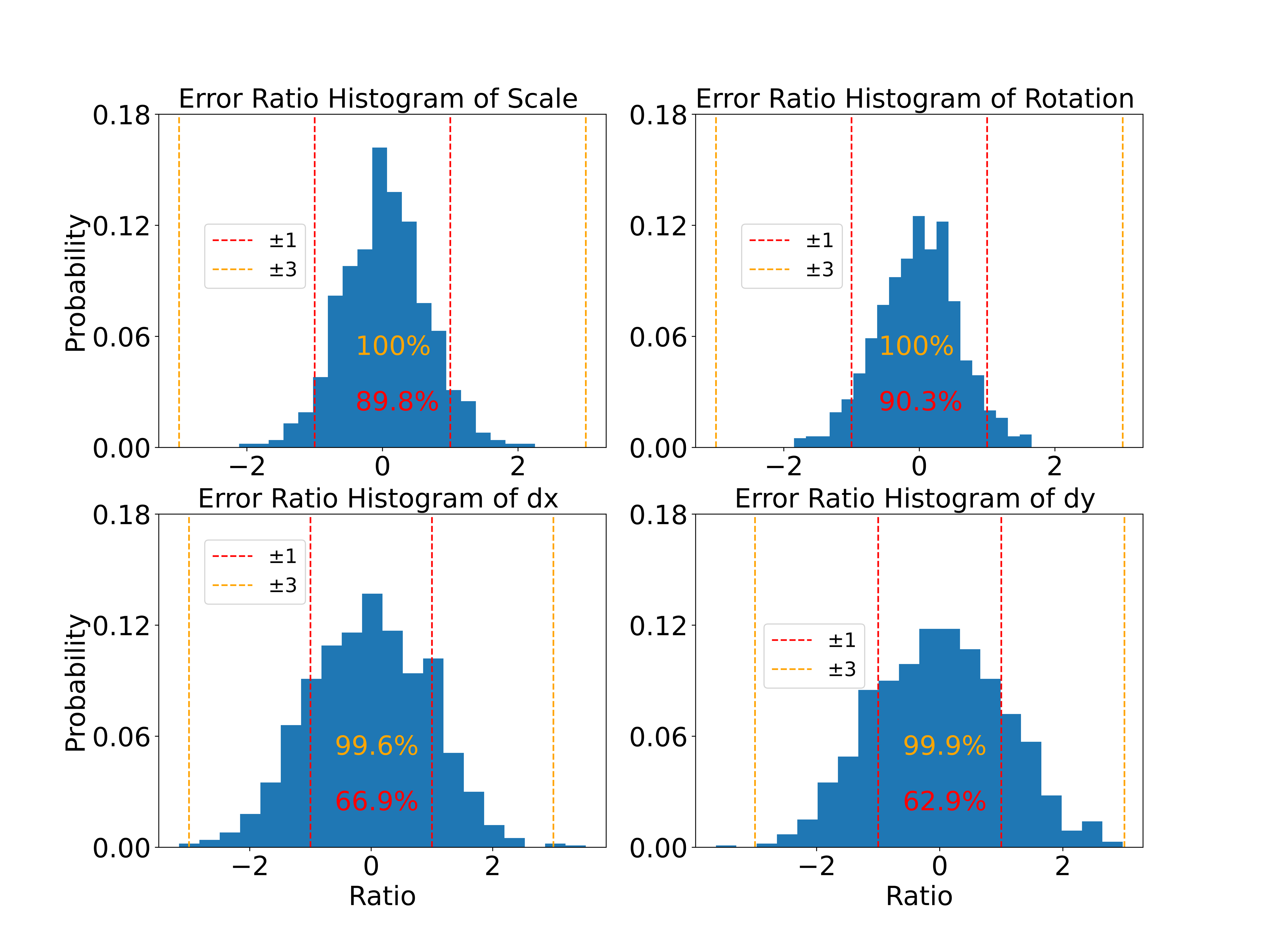}
\caption{\G{Histogram statistics of the error ratios ($s/\sigma$) of the four parameters in AIA 304 {\AA}.} In each measurement, the true residual is denoted by $s$, and the standard error given by the UABJM is denoted by $\sigma$. Inside the red line indicates that the true residual is within 1 times standard error, and inside the yellow line indicates that the true residual is within 3 times standard error. 
\label{fig:err}}
\end{figure}

Before evaluating the accuracy of the method in different wavelengths, the UABJM is performed on the simulated data in Section \ref{subsec:Monte Carlo simulation}. The RMSE of these 1000 groups of standard errors is then utilized as the alignment accuracy (shown in Table \ref{tab:rmse2}). \G{Compared to Table \ref{tab:rmse1}, the accuracies of these two methods are close.}
\G{However, the UABJM is not a statistical method.} It is capable of producing a standard error for each measurement. Then, we need to understand the relationship between the magnitude of the standard error and the true residual in each measurement. In each measurement, the true residual is denoted by $s$, and the standard error given by the UABJM is denoted by $\sigma$. Figure \ref{fig:err} shows the $s/\sigma$ of 1000 measurements represented as a histogram. As can be seen in Figure \ref{fig:err}, the true residuals in the 1000 simulations essentially fall within 3 times the standard error, which is statistically reasonable and valid. Therefore, it is reasonable and effective to use the standard error to evaluate the accuracy of the alignment method on a single wavelength.

\begin{table*}[t]
    \centering
    \caption{RMSE of standard errors in the UABJM.}
    \label{tab:rmse2}
    \begin{tabular*}{\textwidth}{@{\extracolsep{\fill}}lccccr}
    \toprule
        \textbf{RMSE} & \textbf{} & \textbf{Scale} & \textbf{Rotation} & \textbf{x-direction} & \textbf{y-direction} \\ 
        \textbf{} & \textbf{} &  \textbf{$ e^{-6} $} & \textbf{arcsec} & \G{\textbf{pixel}} & \G{\textbf{pixel}} \\
        \midrule
        AIA 304 {\AA} & &  3.9509 & 0.8152 & \G{0.0087} & \G{0.0087} \\ 
        &\G{Add noise} & \G{4.1854} & \G{0.8635} & \G{0.0093} & \G{0.0092} \\
        SDI 1216 {\AA} & & 3.5788 & 0.7384 & \G{0.0090} & \G{0.0088} \\
        &\G{Add noise} & \G{3.5771} & \G{0.7380} & \G{0.0091} & \G{0.0088} \\
        SUTRI 465 {\AA} & &  4.1617 & 0.8585 & \G{0.0046} & {0.0044} \\
        &\G{Add noise} & \G{10.0704} & \G{2.0777} & \G{0.0111} & \G{0.0110} \\
    \bottomrule
    \end{tabular*}
    \tablecomments{Simulation range: scale 0.995:1.005; rotation -0.01:0.01 rad; x-direction -5:5pixel; y-direction -5:5pixel.}
\end{table*}

\subsection{Results} \label{subsec:Results}
Given the reasonable validity of the standard errors derived from the UABJM in single-wavelength simulations, it is feasible to apply this method to estimate the alignment accuracy across different wavelengths. In this experiment, it is applied to alignments of SUTRI 465 {\AA} with AIA 304 {\AA} and SDI 1216 {\AA} with AIA 304 {\AA}. Table \ref{tab:error} shows the standard error estimates for the four parameters for these two alignments.

\G{As illustrated in Table \ref{tab:error}, the standard error estimates of SDI 1216 {\AA} versus AIA 304 {\AA} are larger than those of SUTRI 465 {\AA} versus AIA 304 {\AA} in all aspects. One is because the similarity of the former is smaller than that of the latter; the other is because the difference in spatial resolution of the former is larger. Converting the standard error from arcseconds to pixels based on image size, it can be found that the scaling error of SDI 1216 {\AA} versus AIA 304 {\AA} is 0.5 pixels, the maximum rotation error is 0.77 pixels, and the translation error in both directions is $<$ 0.33 pixels. The alignment accuracy of SUTRI 465 {\AA} and AIA 304 {\AA} is higher. The scaling error of SUTRI 465 {\AA} and AIA 304 {\AA} is 0.08 pixels, the maximum rotation error is 0.12 pixels, and the translation error in both directions is $<$ 0.05 pixels.}

\begin{figure}[ht!]
\plotone{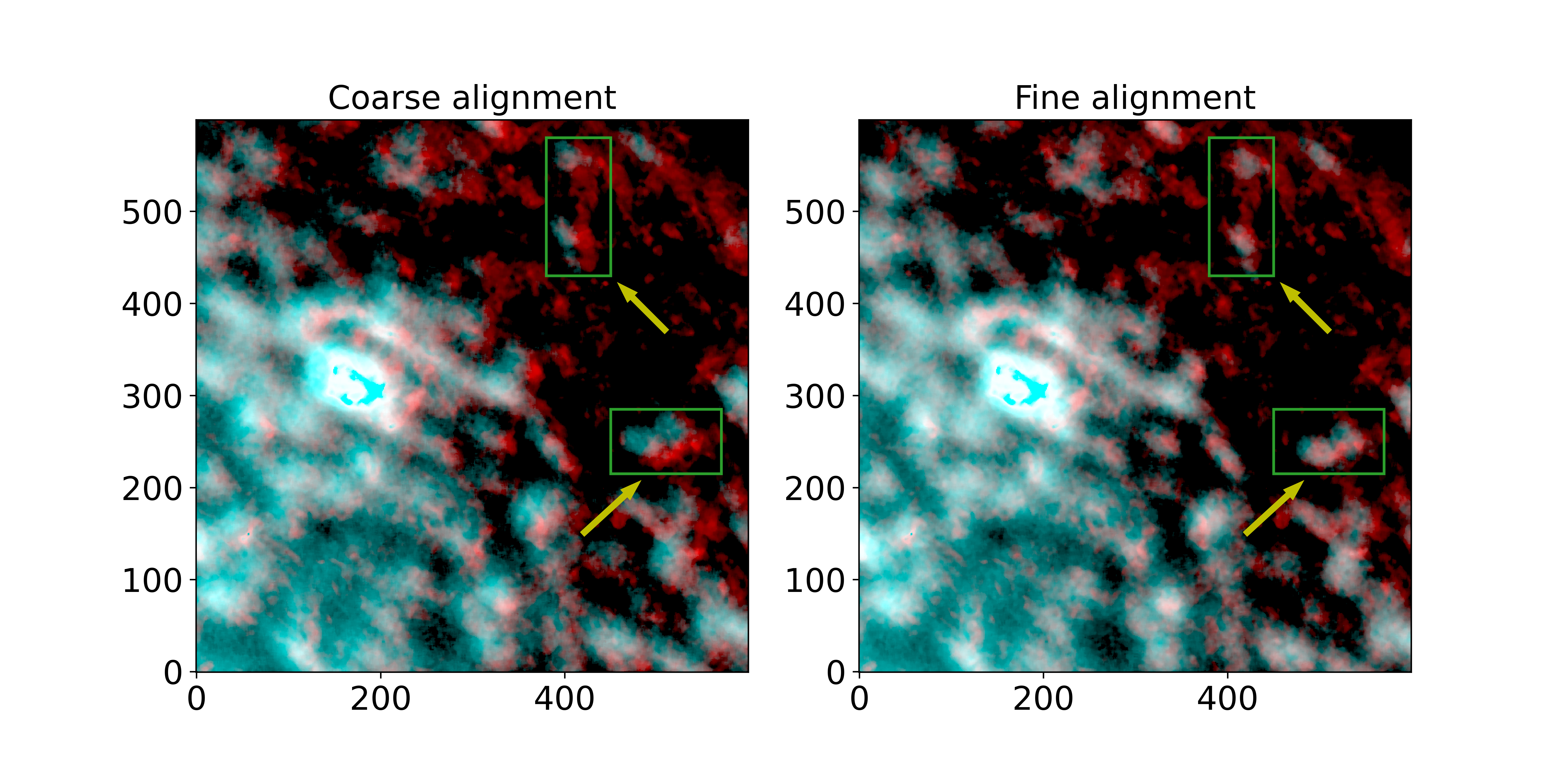}
\caption{SDI 1216 {\AA} and AIA 304 {\AA} alignment results are shown. The left figure shows the coarse alignment results, and the right figure shows the fine alignment. The red channel is AIA 304 {\AA}; the green and blue channels are SDI 1216 {\AA}. The green box is the area with the obvious alignment effect. 
\label{fig:alignment}}
\end{figure}

Among the three groups of data in the experiments in this paper, AIA has the following characteristics: long runtime, stable image quality, and high optical resolution. So we use AIA 304 {\AA} as the alignment standard and align the SUTRI 465 {\AA} image and the SDI 1216 {\AA} image to AIA 304 {\AA}, respectively. To illustrate the alignment results, the two images are synthesized into a pseudo-color composite image. Figure \ref{fig:alignment} shows the alignment results for SDI 1216 {\AA} and AIA 304 {\AA}.\G{ The solar structure overlaps well on the aligned image, and the alignment result is precise.}

\begin{table*}[t]
    \centering
    \caption{Standard errors of alignment for different wavelengths calculated by UABJM in a single measurement}
    \label{tab:error}
    \begin{tabular*}{\textwidth}{@{\extracolsep{\fill}}lccccr}
    \toprule
        \textbf{Stabdard Error} &  \textbf{Scale} & \textbf{Rotation} & \textbf{x-direction} & \textbf{y-direction} & \textbf{image size}\\ 
        \textbf{} &  \textbf{$ e^{-5} $} & \textbf{arcsec} & \textbf{pixel} & \textbf{pixel} & \textbf{pixels$^{2}$}\\
        \midrule
        AIA 304 {\AA}-SDI 1216 {\AA} &  13.52 & 27.68 & 0.28 & 0.32 & 4096$\times$4096\\ 
        AIA 304 {\AA}-SUTRI 465 {\AA} &  4.05 & 8.37 & 0.05 & 0.03  & 2048$\times$2048\\
    \bottomrule
    \end{tabular*}
\end{table*}

\section{Conclusion and Discussion} \label{sec:V.Conclusion and Discussion}
The image alignment of different wavelengths of the sun has historically been a pivotal aspect in the research of the sun. In this paper, a solar image alignment approach is proposed. This approach is based on the dense OF and the RANSAC algorithm. \G{It finds the feature points by extracting the OF fields on these similar structures. From there, it comes to fitting the similarity transformation model to realize the alignment. And the alignment test is performed employing two different groups of images, designated SUTRI 465 {\AA} and AIA 304 {\AA}, along with SDI 1216 {\AA} and AIA 304 {\AA}.} Then, the accuracy of the alignment is evaluated based on the UABJM. \G{The efficacy of the approach is demonstrated through its high degree of accuracy in performing single wavelength simulation alignment experiments, with scale errors $<$ 5$e^{-6}$, rotation errors $<$ 1\(''\), and translation direction errors $<$ 0.01 pixels.} This also indicates that the approach will exhibit high accuracy in the actual alignment work in a single wavelength. And it is expected to be applied in the future to measure the jitter problem of observation instruments. In the actual alignment work in different wavelengths, the accuracy will vary due to the similarity between the images. And the more similar structures, alignment has higher accuracy. We simply use the Bhattacharyya coefficient to quantify the similarity between the images. \G{SUTRI 465 {\AA} and AIA 304 {\AA} exhibits higher similarity.} The alignment of the SDI 1216 {\AA} and AIA 304 {\AA} images can be performed at the sub-pixel level. Furthermore, in the image alignment of SUTRI 465 {\AA} with AIA 304 {\AA}, which exhibits a higher degree of similarity, the pixel error is reduced even further. \G{The rotated pixel errors for each of these alignments are observed to be larger.} In fact, the rotated pixel error on the solar image is not as large as calculated. This is because the calculated error is the maximum rotation pixel error for the image, at the upper right corner of the image (by the lower left corner as the origin). And the fact that the sun is in the center of the image does not include that location.

The alignment approach proposed in this paper is not without its limitations in terms of practical application. Given the OF algorithm's dependence on small motion, this approach is also constrained in its ability to measure large ranges (a significant image position discrepancy) and large motions (such as flares). In the specific code implementation, there exists a pixel window to detect motion. The dimensions of the window, therefore, must be carefully calibrated to ensure that the OF vectors do not exceed their boundaries and fail to identify similar regions. On the other hand, an large window can lead to a decline in the accuracy of the results. Therefore, determining the use of the appropriate window is also a problem, and the topic is not explored in this paper. At the same time, the approach also depends on the degree of similarity between images. When the degree of similarity between two solar images decreases, the accuracy of the approach will be reduced. \G{For example, the degree of similarity between the SDI 1216 {\AA} and AIA 304 {\AA} images in the paper is lower than that between the SUTRI 465 {\AA} and AIA 304 {\AA} images, and the errors of the former are larger than those of the latter.} Furthermore, when the degree of similarity between two solar images is minimal, the alignment approach is rendered ineffective. Therefore, the existence of similar structures between images and small image motion are important prerequisites for the method in this paper.

\G{Our alignment approach is implemented in Python. And the AIA 304 {\AA} and SDI 1216 {\AA} alignment work is publicly available on GitHub: \href{https://github.com/yushiweiliang/Alignment-Method.git.}{https://github.com/yushiweiliang/Alignment-Method.git.}}

\bigskip
We acknowledge the use of data from the Atmospheric Imaging Assembly (AIA) of the Solar Dynamics Observatory (SDO), the Solar Disk Imager (SDI) of the Lyman-alpha (Ly$\alpha$) Solar Telescope (LST), and the Solar Upper Transition Region Imager (SUTRI). We also appreciate all the help from the colleagues in the laboratory team. This work is supported by the National Natural Science Foundation of China under grant 12373115, the Yunnan Key Laboratory of Solar Physics and Space Science under the number 202205AG070009, and the Yunnan Revitalization Talent Support Program under the numbers 202305AS350029 and 202305AT350005.

%




\bibliography{new_article}{}
\bibliographystyle{aasjournal}



\end{document}